\documentclass[11pt]{article}
\usepackage{jcappub}
\usepackage{amsmath,amssymb,amsbsy}
\usepackage{mathtools}

\DeclarePairedDelimiter\abs{\lvert}{\rvert}%
\DeclarePairedDelimiter\norm{\lVert}{\rVert}%
\makeatletter
\let\oldabs\abs
\def\abs{\@ifstar{\oldabs}{\oldabs*}}
\let\oldnorm\norm
\def\norm{\@ifstar{\oldnorm}{\oldnorm*}}
\makeatother

\def\beq{\begin{eqnarray}}
\def\eeq{\end{eqnarray}}

\def\l({\left(}
\def\r){\right)}

\begin{document}


\title{Novel matter coupling in Einstein gravity}

\author{Chunshan Lin\,,\qquad }

\author{Zygmunt Lalak\qquad}

\affiliation{Institute of Theoretical Physics, Faculty of Physics,
University of Warsaw, Ul. Pasteura 5, 02-093 Warsaw, Poland}

\abstract{
A general framework of the novel matter coupling in the Einstein gravity is introduced. We firstly prove that a class of theories whose Hamiltonian constraint is given by an arbitrary function $f(H_g)$, where $H_g$ is the Hamiltonian constraint of general relativity (GR), is equivalent to GR in the vaccum. A novel Jordan frame is defined when GR is rewritte in terms of one of its equivalents in this class. The transformation between the novel Jordan frame and the Eintein frame is a redifinition of lapse.  We discuss two types of consistency condition for matter to couple to Einstein gravity in the novel Jordan frame. The Type I consistency condition is found by demanding all constraints to be first class; the type II consistency condition is found by demanding the algebra is closed when matter minimally couples to gravity in the novel Jordan frame, which an additional gauge condition is required. We discuss the cosmological implications from these two types of matter coupling. 
}

\maketitle

\section{Introduction}

General relativity (GR) is an extremely simple and elegant theory that describes all known gravitational interactions in 4 dimensional space-time.   Its underlying symmetry is the space-time diffeomorphism invariance, which leads to the equivalence principle in the low energy limit. It is precisely the equivalence principle that Einstein began with, embarking himself on what would be an eight-year search for a relativistic theory of gravity. Formulated in the framework of Riemann geometry,  the space-time is viewed as a Riemannian manifold, and gravity is interpreted as a geometric property of space-time, rather than an external  force known from the Newtonian theory of gravity. The space-time diffeomorphism invariance is, naively speaking, a local gauged version of space-time translation invariance under which all  physical variables are transformed.  This symmetry must be preserved as we introduce matter content to be coupled to gravity, and the Noether theorem warrants us a conserved energy momentum tensor. The minimal coupling and the Brans–Dicke type of coupling are two well known examples of this kind which manifestly respects the space-time diffeomorphism. It is thus very intriguing to ask whether there exists any new type of matter coupling which also respects the diffeomorphism, but may not in a manifest manner? This is the question that we are trying to address in our current work.

At the action level, the space-time diffeomorphism invariance is manifest when all tensors and vectors contract with metric tensor $g^{\mu\nu}$ and form scalars. While in the Hamiltonian, the manifestation of the space-time diffeomorphism invariance is exhibited in terms of 8 first class constraints  \cite{dirac1}\cite{dirac2}, which  eliminate 16 degrees in the phase space and, as a consequence, graviton is massless and it has only two polarizations in its spectrum. 
A self-consistent matter coupling must maintain this appealing algebraic structure of the theory.  However, the formulation of GR in the Hamiltonian language is not unique, and one form can be mapped onto another via canonical transformation  \cite{Aoki:2018zcv}\cite{Aoki:2018brq}. Therefore, a more basic question that one should ask, prior to the one of novel matter coupling, is whether GR is the unique theory of which all constraints are first class? Or in other words, is there any other theory which is as good as GR in the sense that all constraints are first class, and therefore the structure of the theory is protected by these local gauge symmetries associated to  the first class constraints? This question was initially formulated in Ref. \cite{Lin:2017oow}, in which the first example that seemly differs from GR, i.e. the so called square-root gravity, was discovered by solving the consistency condition that ensures a closed algebra and all constraints to be first class.  

This framewhork was later extended to the whole class of theories of which the Hamiltonian constraint is written as an arbitrary function $f(H_g)$, where $H_g$ is the Hamiltonian constraint of GR \cite{Carballo-Rubio:2018czn} (see also Ref. \cite{Mukohyama:2019unx}). On the other hand, the graviton scattering amplitude exhibits a perturbative equivalence between the square-root gravity and GR which holds up to 5 point function level for all possible helicity configurations in the Minkowskian vacuum. This perturbative equivalence  delivers two important messages: (1) this whole class of theories is probably just GR in a different guise, which we will explicitly prove by two independent non-perturbative approaches  in our current paper; (2) the minimal coupling of the square-root gravity renders the algebra unclosed \cite{Carballo-Rubio:2018czn}, while a novel matter coupling is required to close it \cite{Lin:2018mip}. This implies the existence of a systematic and general framework to couple matter to gravity in a self-consistent manner, while it offers us much richer phenomenology. It is the main objective of our current work to find and  study this general framework.  

This paper is organized as follows: we will introduce a class of GR equivalents in the section \ref{GReq}. We will discuss how to couple matter to gravity in the self-consistent manner in the section \ref{matcoup}. Phenomenology is discussed in the section \ref{pheno}. We conclude in the section \ref{conclusion}.

\section{A class of GR equivalents}\label{GReq}
In this section, we introduce a class of GR equivalents, whose equivalence to the Einstein gravity is proved by means of two non-perturbative approaches, the Hamiltonian analysis a la Dirac \cite{dirac1}\cite{dirac2}, and the equation of motion. We adopt the ADM decomposition, in which the space-time metric reads
\beq
ds^2=-N^2dt^2+h_{ij}\left(dx^i+N^idt\right)\left(dx^j+N^jdt\right),
\eeq
where $h_{ij}$ is the 3-dimensional induced metric, $N$ is the lapse and $N^i$ is the shift. 

\subsection{Hamiltonian analysis}
This class of theories, whose first example was discovered in Ref. \cite{Lin:2017oow} and later was extended in Ref. \cite{Carballo-Rubio:2018czn} (see also \cite{Mukohyama:2019unx}), contains 8 constraints in the Hamiltonian, where 4 of them are primary ones and another 4 of them are secondary ones. All these 8 constraints are first class, i.e.  Poisson brackets between any two constraints vanish on the constraint surface. To determine the nature of this class of theories, one of the most direct methods is to find the local gauge symmetry generators associated with all first class constraints. To this end, we adopt the Hamiltonian analysis approach introduced by Dirac \cite{dirac1}\cite{dirac2}, and write down the following Hamiltonian as our starting point,
\beq\label{eq2gr}
H=\int d^3x\left(N\mathcal{H}_0+N^i\mathcal{H}_i+\lambda_N\pi_N+\lambda^i\pi_i\right),
\eeq
where $\pi_N$ is the conjugate momentum of the lapse $N$, $\pi_i$ is the conjugate momentum of the shift $N^i$, $\lambda_N$ and $\lambda^i$ are Lagrangian multipliers that enforce the following 4 primary constraints,
\beq
\pi_N\approx0,\qquad\pi_i\approx 0\,.
\eeq
These 4 primary constraints must be preserved by time evolution of the system. The consistency conditions then give us another 4 secondary constraints, they are the Hamiltonian constraint and the momentum constraints,
\beq
\mathcal{H}_0\approx0,\qquad\mathcal{H}_i\approx0\,,
\eeq
where $\mathcal{H}_i\approx0$ are the momentum constraints,
\beq
\mathcal{H}_i &\equiv&  -2 \sqrt{h}\nabla_j\left(\frac{\pi^j_{~i}}{\sqrt{h}}\right)\,,
\eeq
where $h_{ij}$ is the induced 3-metric, $\pi^{ij}$ is the conjugate momentum of $h_{ij}$, and $\nabla_i$ is the covariant derivative compatible to $h_{ij}$, and the $\mathcal{H}_0\approx0$ is the Hamiltonian constraint, it can be written as a generic function $f(H_g) $ of its argument,
\beq\label{H0}
\mathcal{H}_0=\sqrt{h}f\left(H_g\right), \qquad H_g\equiv R+\frac{\lambda}{h}\left(\pi^{ij}\pi_{ij}-\frac{1}{2}\pi^2\right),
\eeq
where  $R$ is the Ricci scalar of 3-d induced metric. Noted that we have only considered the pure gravity in a vacuum in this section, the inclusion of matter sector will be discussed later. The $\lambda$ in the eq.(\ref{H0}) is an arbitrary constant, it varies if we rescale the space-time coordinates.  In the case of GR, we have $\mathcal{H}_0= -H_g/2$ and we set $\lambda=-4$ conventionally (so that the speed of light is unity). In the square-root gravity, we have $\mathcal{H}_0\sim\sqrt{H_g+\Lambda_1}+\Lambda_2$, where $\Lambda_1$ and $\Lambda_2$ are constants. The spatial diffeomorphism invariance is manifest preserved, while the temporal diffeomorphism seems explicitly broken. 

Now let us compute all Poisson brackets.  In our paper, we adopt the following useful notations,
\beq
\mathcal{O}[\alpha]\equiv\int d^3x\alpha\mathcal{O},\qquad\mathcal{O}_i[f^i]\equiv\int d^3xf^i \mathcal{O}_i.
\eeq
 According with the previous results \cite{Carballo-Rubio:2018czn}, all brackets are vanishing weakly on the constraint surface, 
\beq
\{\mathcal{H}_0[\alpha],\mathcal{H}_0[\beta]\}&\approx& 0,\qquad \{\mathcal{H}_0[\alpha],\mathcal{H}_i[f^i]\}\approx 0,\qquad \{\mathcal{H}_0[\alpha],\pi_N[\beta]\}\approx 0,\nonumber\\
\{\mathcal{H}_0[\alpha],\pi_i[f^i]\}&\approx& 0,\qquad  \{\mathcal{H}_i[f^i],\mathcal{H}_i[g^i]\}\approx 0,\qquad  \{\mathcal{H}_i[f^i],\pi_N[\alpha]\}\approx 0, \nonumber\\
 \{\mathcal{H}_i[f^i],\pi_i[g^i]\}&\approx& 0,\qquad   \{\pi_N[\alpha],\pi_N[\beta]\}\approx 0,\qquad   \{\pi_N[\alpha],\pi_i[f^i]\}\approx 0,\nonumber\\
   \{\pi_i[f^i],\pi_N[g^i]\}&\approx & 0,\qquad 
\eeq
where $\alpha$, $\beta$, $f^i$ and $g^i$ are arbitrary functions that depend on space and time. The algebra closes and therefore all constraints are first class. It implies the existence of a mysterious local gauge symmetry, in addition to the spatial diffeomorphism invariance, prohibits the longitudinal mode of graviton. It turns out that this mysterious local gauge symmetry is nothing but temporal diffeomorphism, which we will  prove it in the rest of this subsection. 

It is indeed somewhat confusing since the action of this class of theories, which can be obtained via a Legendre transformation, does not look manifestly general covariant, but it is actually equivalent to a theory that fully respects all space-time diffeomorphism, i.e. the Einstein gravity. There is actually a similar example in the literature, where the Einstein gravity is reduced to the BSW action  by integrating out the lapse \cite{shape1}, which also seemly breaks temporal diffeomorphism. Nevertheless, the Einstein gravity and the BSW action are completely equivalent. In our case, one of the important evidences of the equivalence is that the Hamiltonian constraint and the momentum constraints serve as generators of the space-time diffeomorphism.  Noted that our theories are written in the manifestly spatial diffeomorphism invariant manner, thus the momentum constraints simply generate the spatial diffeomorphism,
\beq\label{mgen}
\{h_{ij},\mathcal{H}_i[f^j]\}=\nabla_if_j+\nabla_jf_i.
\eeq
Now let's check the Hamiltonian constraint. Firstly, the Hamiltonian's equation of motion gives us
\beq\label{eomhij}
\dot{h}_{ij}=\{h_{ij}, H\}\approx  \frac{\partial f}{\partial H_g}\frac{2\lambda N}{\sqrt{h}}\left(\pi_{ij}-\frac{1}{2}\pi h_{ij}\right)+\nabla_iN_j+\nabla_jN_i, 
\eeq
the conjugate momentum evaluates to 
\beq\label{momh}
\lambda\frac{\partial f}{\partial H_g}\frac{\pi_{ij}}{\sqrt{h}}=K_{ij}-Kh_{ij},
\eeq
where $K_{ij}$ is the extrinsic curvature $K_{ij}\equiv \frac{1}{2N}\left(\partial_t h_{ij}-\nabla_iN_j-\nabla_jN_i\right)$. The conjugate momentum, and thus the gravity theory, become ill defined if $\frac{\partial f}{\partial H_g}$ vanishes weakly on the constraint surface. Therefore, throughout this paper, we only focus on the theories whose $\frac{\partial f}{\partial H_g}$ is not vanishing (not even weakly!). 
The Hamiltonian constraint, as a first class constraint, generates the following local gauge transformation, 
\beq\label{Hgen}
\{h_{ij}, \mathcal{H}_0[\xi]\}\approx \frac{\partial f}{\partial H_g}\frac{2\lambda\xi}{\sqrt{h}}\left(\pi_{ij}-\frac{1}{2}\pi h_{ij}\right)=\xi\partial_th_{ij}-\xi\partial_jN_i-\xi\nabla_iN_j,
\eeq
where the eq. (\ref{momh}) has been used, we have absorbed the lapse into the redefinition of $\xi$, and $N_i\equiv N^j h_{ij}$.   The temporal diffeomorphism generator is the combination of the Hamiltonian constraint and the momentum constraints, i.e.
\beq
\mathcal{T}[\xi]\equiv \int \xi \left(\mathcal{H}_0+ N^i\mathcal{H}_i \right)d^3x. 
\eeq
We can check that it does generate the temporal diffeomorphism $t\to t+\xi\left(t,\textbf{x}\right)$,
\beq
\{h_{ij}, \mathcal{T}[\xi]\}\approx\xi\partial_th_{ij}+N_i\partial_j\xi+N_j\partial_i\xi=\pounds_t h_{ij}. 
\eeq

Therefore, this class of theories whose Hamiltonian is written in the form of eq. (\ref{eq2gr}) is equivalent to the GR in a vacuum, in the sense that both of local gauge symmetries are the space-time diffeomorphism invariance and thus the physical observables are unaffected under the diffeomorphism. However, we have to emphasize that it is not yet clear which transformation maps the GR to the theories of  eq. (\ref{eq2gr}). 

\subsection{The equation of motion}
The equation of motion of graviton is another perspective from which we can see the equivalence between GR and the class of theories written in the eq. (\ref{eq2gr}).  In this subsection, we will derive the equation of motion and explicitly show this equivalence. Given the Hamiltonian eq. (\ref{eq2gr}), the Hamilonitan's equation of motion is given by the eq. (\ref{eomhij}), and the conjugate momentum of the induced metric $h_{ij}$ is given by the eq. (\ref{momh}).  After a straightforward computation, we get the equation of motion for graviton,
\beq\label{eom_vacuum}
\frac{d}{dt}\left[\left(\frac{\partial f}{\partial H_g}\right)^{-1}K_{ij}\right]&=&\{\frac{\lambda}{\sqrt{h}}\left(\pi_{ij}-\frac{1}{2}\pi h_{ij}\right),H\}\nonumber\\
&=&\left(\frac{\partial f}{\partial H_g}\right)^{-1}\left[2NK^k_{~i}K_{kj}-NKK_{ij}-\frac{1}{2}Nh_{ij}\left(K^{kl}K_{kl}-K^2\right)+\pounds_NK_{ij}\right]\nonumber\\
&&+\lambda \frac{\partial f}{\partial H_g}\left(NG_{ij}-\nabla_i\nabla_j N\right)\nonumber\\
&&-\lambda N\nabla_i\nabla_j\left(\frac{\partial f}{\partial H_g}\right)+K_{ij}N^k\nabla_k\left(\frac{\partial f}{\partial H_g}\right)^{-1},
\eeq
where the Lie derivative is directed along the shift $N^i$, i.e. 
\beq
\pounds_NK_{ij}=N^l\nabla_lK_{ij}+K^l_{~i}\nabla_lN_j+K^l_{~j}\nabla_lN_i.
\eeq	
The Hamiltonian constraint is an algebraic equation. Assuming that it has at least one solution,  we have then
\beq\label{onshell}
f\left(H_g\right)\approx0\qquad\to \qquad H_g=\text{constant}\qquad\to\qquad\frac{\partial f}{\partial H_g}=\text{constant},
\eeq
and therefore those two terms at the last line of eq. (\ref{eom_vacuum}) vanish, and the equation of motion simplifies to 
\beq
\frac{d}{dt} K_{ij}&=&2NK^k_{~i}K_{kj}-NKK_{ij}+\pounds_NK_{ij}\nonumber\\
&&+\lambda \left(\frac{\partial f}{\partial H_g}\right)^2\left(NR_{ij}-\frac{1}{2}N\Lambda h_{ij}-\nabla_i\nabla_j N\right),
\eeq
where we have used the Hamiltonian constraint $K^{ij}K_{ij}-K^2=\lambda\left(\frac{\partial f}{\partial H_g}\right)^2\left(\Lambda-R\right)$ to simplify the above equation. 
Noted that the factor $ \left(\frac{\partial f}{\partial H_g}\right)^2$ at the last line of the above equation can be absorbed into a rescaling of $\lambda$ (which amounts to a space-time coordinate rescaling), the factor $\frac{\partial f}{\partial H_g}$ thus drops out of the equation of motion. At the end of the day, the equation of motion for graviton coincides with the one in GR, regardless of the form of $f(H_g)$ (as long as Hamiltonian constraint has at least one real solution). On the other hand, from eq. (\ref{onshell}) it is easy to see that the Hamiltonian constraint and the momentum constraints coincide with the ones of GR too. We conclude that this class of theories is equivalent to the Einstein gravity also at the equation of motion level, as it should be since the equation of motion is invariant under the space-time diffeomorphism, which has shown to be the local gauge symmetry of the theories in the preceding subsection. \\

It is quite remarkable that the original motivation, which eventually leads to the discovery of this class of GR equivalents,  is to look for the theories which are as good as GR in the sense that all constraints are first class \cite{Lin:2017oow}. However, it has turned out that in 4 dimensional space-time, the only theory  we have found  is the GR itself. Nevertheless, we would like to mention that this unique and distinctive role of GR is still challengeable, on the account of modifying the action principle \cite{Glavan:2019inb}.

\section{The self-consistent matter coupling of the GR equivalents}\label{matcoup}
In the last section, we have demonstrated  that a class of theories whose Hamiltonian is written in the form of eq. (\ref{eq2gr}) is equivalent to GR. Rewriting GR in terms of one of its equivalents, we actually define a new frame in which all physical laws are embedded. It is natural to ask how to couple matter to gravity in the new frame. The simplest one could be the minimal coupling between gravity and matter in the new frame. However, it has turned out that this type of coupling renders the algebra unclosed and the theory becomes inconsistent  \cite{Carballo-Rubio:2018czn}. In this section, we will develop a systematic and general framework in which matter can couple to gravity in the theoretically self-consistent manner. We will start from a single scalar field, as it is the simplest, and then generalize it to the multi-field as well as the higher spins. 

\subsection{Type I consistency condition}
The type I consistency condition ensures the closure of algebra close and that all constraints  to be first class. Let's start from the simplest single scalar field. Assuming that the spatial diffeomorphism is still manifestly invariant and thus we have the momentum constraints written as
\beq\label{mc}
\mathcal{H}_i=-2\sqrt{h}\nabla_j\left(\frac{\pi^j_{~i}}{\sqrt{h}}\right)+\pi_{\phi}\nabla_i\phi\approx0,
\eeq
which also serves as the spatial diffeomorphism generators for both of graviton $h_{ij}$ and the scalar $\phi$. 
Inspired by Ref. \cite{Lin:2018mip}, we adopt the ansatz that the Hamiltonian constraint is written as an algebraic function of its arguments
\beq\label{H0gm}
\mathcal{H}_0=\sqrt{h}f\left(H_g,H_m\right)\approx0, 
\eeq
where $H_g$ is defined in the eq. (\ref{H0}) and $H_m$ is the would-be Hamiltonian of the scalar field if minimally coupled with the Einstein gravity in the Einstein frame,
\beq
H_m\equiv\zeta_1\frac{\pi_{\phi}^2}{h}+\zeta_2\nabla_i\phi\nabla^i\phi+\zeta_3,\qquad\zeta_i\equiv\zeta_i(\phi),
\eeq
where $\zeta_1$, $\zeta_2$ and $\zeta_3$ are three arbitrary functions of the scalar field $\phi$. In our general setup, the Hamiltonian constraint is a generic function of its arguments $H_g$ and $H_m$. 
We work out all Poisson brackets in the following: 
\beq
\{\mathcal{H}_0[\alpha],\mathcal{H}_i[f^i]\}&\approx& 0,\qquad \{\mathcal{H}_0[\alpha],\pi_N[\beta]\}\approx 0,\qquad\{\mathcal{H}_0[\alpha],\pi_i[f^i]\}\approx 0,\nonumber\\
 \{\mathcal{H}_i[f^i],\mathcal{H}_i[g^i]\}&\approx& 0,\qquad  \{\mathcal{H}_i[f^i],\pi_N[\alpha]\}\approx 0, \qquad  \{\mathcal{H}_i[f^i],\pi_i[g^i]\}\approx 0,\nonumber\\
  \{\pi_N[\alpha],\pi_N[\beta]\}&\approx& 0,\qquad   \{\pi_N[\alpha],\pi_i[f^i]\}\approx 0,\qquad    \{\pi_i[f^i],\pi_N[g^i]\}\approx 0,
\eeq
and
\beq
\{\mathcal{H}_0[\alpha],\mathcal{H}_0[\beta]\}=-2\int \left(\beta\nabla^i\alpha-\alpha\nabla^i\beta\right)\left[\lambda\left(\frac{\partial f}{\partial H_g}\right)^2\cdot\sqrt{h}\nabla^i\left(\frac{\pi_{ij}}{\sqrt{h}}\right)+2\zeta_1\zeta_2\left(\frac{\partial f}{\partial H_m}\right)^2\pi_\phi\nabla_i\phi\right].\nonumber\\
\eeq
Therefore, the only non-trivial Poisson bracket is the one with Hamiltonian constraint commuting with itself, which vanishes weakly on the constraint surface if 
\beq\label{cc0}
\lambda\left(\frac{\partial f}{\partial H_g}\right)^2\approx-4\cdot\zeta_1\zeta_2\cdot\left(\frac{\partial f}{\partial H_m}\right)^2.
\eeq
Noted that the product $\zeta_1\cdot\zeta_2$ must be a constant, otherwise the above self-consistency condition can never be satisfied, given the Hamiltonian constraint in eq. (\ref{H0gm}). Throughout this paper, we fix their relation with $\lambda$ so that  $\lambda=-4\zeta_1\zeta_2$. This relation can always be realized via either the space-time coordinate rescaling or the variable redefinition $f(H_m)\to \tilde{f}\left(\tilde{H}_m\right)$, where $\tilde{H}_m\equiv \text{constant}\cdot H_m$. The condition eq. (\ref{cc0}) thus acquires a  dramatically simple form,
\beq\label{cc}
\left(\frac{\partial f}{\partial H_g}\right)^2\approx\left(\frac{\partial f}{\partial H_m}\right)^2.
\eeq
This is the type I consistency condition of the matter coupling in the GR equivalents, which ensures that all constraints to be first class. 

The type I consistency condition seemly grants innumerous solutions, however, almost all of solutions are equivalent to the minimal coupling in general relativity. To see this, let's alter both of $H_g$ and $H_m$ by small values $\delta H_g$ and $\delta H_m$. On the constraint surface, we have 
\beq\label{deltaf}
\delta f(H_g,H_m)=\frac{\partial f}{\partial H_g}\delta H_g +\frac{\partial f}{\partial H_m}\delta H_m=0,
\eeq
where $\delta f=0$ is to ensure that we are still on the constraint surface after altering the values of $H_g$ and $H_m$. Combining this equation with the type I consistency condition eq. (\ref{cc}), immediately we can see that $\delta H_g=\pm \delta H_m$, namely $H_g$ must be linear in $H_m$ for the solution to Hamiltonian constraint \footnote{We are not interested in the case that either $\partial f/\partial H_g$ or $\partial f/\partial H_m$ is vanishing.}, which coincides with the solution of minimal coupling in general relativity. 

The same analysis can be generalized to the case of multi scalar fields. We again assume that the theory is manifestly spatial diffeomorphism invariant, and the momentum constraint is written as 
\beq
\mathcal{H}_i=-2\sqrt{h}\nabla_j\left(\frac{\pi^j_{~i}}{\sqrt{h}}\right)+\sum_I\pi_{I}\nabla_i\phi^I\approx0,
\eeq
where $\pi_I$ is the conjugate momentum of the scalar $\phi^I$. The consistency condition eq. (\ref{cc}) is accordingly extended to 
\beq\label{cc2}
\left(\frac{\partial f}{\partial H_g}\right)^2\approx\left(\frac{\partial f}{\partial H_m^I}\right)^2\approx\left(\frac{\partial f}{\partial H_m^J}\right)^2,
\eeq
where 
\beq
H_m^I\equiv\zeta_1^I\frac{\pi_{I}^2}{h}+\zeta_2^I\nabla_i\phi^I\nabla^i\phi^I+\zeta_3^I, \qquad -4\zeta_1^I\zeta_2^I=\lambda. 
\eeq
Once again, $H_g$ must be linear in $H_m^I$ for the solution of Hamiltonian constraint, given the consistency condition eq. (\ref{cc2}).

Let's give an example of this kind. We adopt the ansatz that both of gravity Hamiltonian and matter Hamiltonian can be written in the following power law form,
\beq\label{ex1}
\mathcal{H}_0=\xi\sqrt{h} \left[\left(\Lambda_1H_g+\Lambda_2\right)^n-\left(\Lambda_3H_m+\Lambda_4\right)^p\right]\approx0,
\eeq
where $ \xi=\pm1$ and $\Lambda_i's$ are constants. The consistency condition eq. (\ref{cc}) implies 
\beq
n=p, \qquad\Lambda_1^2=\Lambda_3^2.
\eeq
The Hamiltonian constraint can be rewritten in terms of the product between its solution $H_g=\pm H_m+\text{constant}$ and a non-vanishing factor that can be aborbed into the redefinition of lapse. For instance, for $n=2$, $\xi=-1$ and $\Lambda_1>0$, the Hamiltonian reads
\beq
H=\int d^3x\sqrt{h} \tilde{N}\left(-H_g+H_m-\frac{\Lambda_2-\Lambda_4}{\Lambda_1}\right)+N^i \mathcal{H}_i,
\eeq
where 
\beq
\tilde{N} \equiv N \Lambda_1^2\left(H_g+H_m+\frac{\Lambda_2+\Lambda_4}{\Lambda_1}\right)
\eeq
is the extended lapse. This is precisely the Hamiltonian of a canonical scalar field that minimally couples to general relativity.

Nevertheless, an exception (probably the only one) does exist, where
\beq\label{type1excep}
H_0= \sqrt{h}\left[-H_g+\abs{H_m}+\Lambda\right],
\eeq
which also renders the closure of algebra, and all contraints are first class. One of the appealing features of this theory is that the matter Hamiltonian is always bounded from below, at the expense of losing the continuity around $H_m=0$, which may grant us some interesting cosmological and astrophysical applications. We will study the cosmology of this theory in the next section.


\subsection{Type II consistency condition}
As we have learned in the preceding section, we have defined a new frame by rewriting general relativity in terms of its equivalents. Previous studies suggested that the simplest matter coupling, i.e. minimal coupling  in the new frame gives rise to the odd dimensionality of phase space, and thus the system is pathological \cite{Carballo-Rubio:2018czn}. The minimal coupling in the new frame corresponds to a novel matter coupling which partially breaks the temporal diffeomorphism. To close the algebra, we need to introduce a constraint by hand to fix the leftover gauge freedom \cite{Aoki:2018zcv}\cite{Aoki:2018brq}\cite{Mukohyama:2019unx}, namely, 
\beq\label{type2cond}
H=\int d^3x\left\{ \sqrt{h}N\left[f\left(H_g\right)+H_m\right]+N^i\mathcal{H}_i+\lambda\mathcal{G}\right\},
\eeq
where $\lambda$ is a Lagrangian multiplier and the gauge condition $\mathcal{G}\approx0$ is a 3d scalar function of the canonical variables. The Hamiltonian constarint $\mathcal{H}_0=f(H_g)+H_m\approx0$ and $\mathcal{G}\approx0$ become a pair of second class constraints, eleminate a pair of degrees in the phase space.

Let's offer an example of type II matter coupling, where the Hamiltonian in the Jordan frame is written by
\beq\label{type2hj}
H_J=\int d^3x\left[\sqrt{h}N\left(-\frac{H_g^2}{4M^4}+H_m\right)+N^i\mathcal{H}_i+\lambda\mathcal{G}\right].
\eeq 
The Hamiltonian in the Einstein frame can be obtained by rescaling lapse
\beq
N&\to& NM^4/\left(\frac{1}{2}H_g+M^2\sqrt{H_m}\right),\\
H_E&=&\int d^3x\left[\sqrt{h}N\left(-\frac{1}{2}H_g+M^2\sqrt{H_m}\right)+N^i\mathcal{H}_i+\lambda\mathcal{G}\right],\label{type2he}
\eeq
We then take the Legendre transformation in order to
obtain the Lagrangian corresponding to the Hamiltonian eq. (\ref{type2he}). For this purpose, we redefine the Lagrange multiplier and the gauge condition as
\beq
\lambda\to \lambda N,\qquad \mathcal{G}\to\sqrt{h}\mathcal{G}.
\eeq
Just for simplicity, let's assume that the gauge condition $\mathcal{G}$ does not contain $\pi^{ij}$ or $\pi_{\phi}$. We obtain
\beq
\dot{h}_{ij}=\{h,H\}=\frac{4 N}{\sqrt{h}}\left(\pi_{ij}-\frac{1}{2}\pi h_{ij}\right)+\nabla_iN_j+\nabla_jN_i,
\eeq
and then the conjugate momentum reads
\beq
\pi_{ij}=\frac{1}{2}\sqrt{h}\left(K_{ij}-Kh_{ij}\right).
\eeq
On the other hand, in the scalar sector we have 
\beq
\dot{\phi}=\{\phi,H\}=\frac{NM^2}{\sqrt{H_m}}\frac{\pi_{\phi}}{\sqrt{h}}.
\eeq
After the Legendre transformation, the action and the Lagrangian reads
\beq\label{sqrphi}
S&=&\int d^4x\sqrt{-g}\mathcal{L},\nonumber\\
\mathcal{L}&=&\frac{1}{2}M_p^2\mathcal{R}-\sqrt{\left[M^4-\frac{1}{N^2}\left(\dot{\phi}-N^i\partial_i\phi\right)^2\right]\left(h^{ij}\nabla_i\phi\nabla_j\phi+V(\phi)\right)}-\lambda\mathcal{G}.
\eeq



\section{Phenomenologies}\label{pheno}
The novel matter coupling discovered in our current work opens up new possibilities for phenomenological studies. In this section, we will explore some aspects of its phenomenologies, including the spherical static solution and FLRW cosmologies. 

\subsection{Schwarzschild solution}\label{Schwarz}
We now derive the static spherically symmetric solution induced by a point mass, as it is one of the most basic phenomenologies.   Let's take a spherical static ansatz, 
\beq\label{sssans}
ds^2=-A(r)dt^2+B(r)dr^2+r^2\left(d\theta^2+\sin^2\theta d\varphi^2\right).
\eeq
The Hamiltonian of gravity is given in the eq. (\ref{eq2gr}), and the  Einstein tensor are derived in the appendix \ref{MET}.  Under the spherical static ansatz, the Einstein tensor reduces to 
\beq\label{etensorSSS}
G^0_{~0}&=&-f,\qquad G^0_{~i}=0,\qquad G^{i}_{0}=0,\nonumber\\
G^i_{~j}&=&-\delta^i_{~j}f+2\frac{\partial f}{\partial H_g}R^{i}_{~j}-\frac{2}{N}\left(\nabla^i\nabla_j-\delta^i_{~j}\Delta\right)\left(N\frac{\partial f}{\partial H_g}\right).
\eeq
The $00$ component of the Einstein equation leads to
\beq
f\left(H_g\right)=0.
\eeq
it implies that 
\beq
H_g\equiv R+\frac{\lambda}{h}\left(\pi^{ij}\pi_{ij}-\frac{1}{2}\pi^2\right)=\text{constant}. 
\eeq
The constant in the above equation can be canceled out by a bare cosmological constant and we have a Minkowskian solution in the absence of mass. With a point mass included, the static ansztz eq. (\ref{sssans}) implies the conjugate momentum $\pi^{ij}$ of 3-d induced metric is vanishing and we have 
\beq
R=0,\qquad\to\qquad B(r)=\left(1-\frac{r_S}{r}\right)^{-1},
\eeq
where the integral constant  $r_S=2GM$ is fixed by the boundary condition at infinity large distance $B-1\to \frac{2GM}{r}$ as $r\to \infty$. Noted that we also have $\frac{\partial f}{\partial H_g}=\text{constant}$ as its argument $H_g$ is a constant, and thus the trace of the $ij$ component of the Einstein tesnor eq. (\ref{etensorSSS}) implies 
\beq
\partial_r\left(r^2\partial_rA\right)=0, 
\eeq
and again we have 
\beq
A(r)=1-\frac{r_S}{r},
\eeq
and $r_S=2GM$ is fixed by the boundary condition at infinity large distance $A-1\to -\frac{2GM}{r}$ as $r\to \infty$.

It is not surprising that the spherical static vacuum solution coincides with the Schwarzschild solution, because our theory is equivalent to GR in the vacuum. However, new predictions would appear if matter couples to our theory in the novel manner introduced in the last section.

\subsection{FLRW cosmology}
We will discuss the cosmological solutions arising from the theories eq. (\ref{type1excep}) and eq. (\ref{type2cond}) respectively in this section. 

\subsubsection{a toy model for the type I novel coupling}
The only nontrivial solution to the self-consistency condition eq. (\ref{cc}) is the theory eq. (\ref{type1excep}), where the gravity Hamiltonian is the one of GR, while the matter Hamiltonian is positive defined and thus always bounded from below. In principle it allows us to turn on a ghost to achieve an exotic phase transition, for instance an null energy condition (NEC) violation phase such as cosmic bounce, and turn it off after the transition. We would like to propose a toy model to illustrate this idea, where in a contracting universe the Hamiltonian constraint reads
\beq
\mathcal{H}_0=\frac{1}{2}\sqrt{h}\left\{\Lambda_1-H_g+H_{\text{radiation}}+\frac{\pi_\phi^2}{h}+\nabla^i\nabla_i\phi-m^2\phi^2\right.\nonumber\\
\left.+\abs{\frac{\pi_\chi^2}{h}+\nabla_i\chi\nabla^i\chi+\lambda\phi^4-\Lambda_2}\right\},
\eeq
where $H_{\text{radiation}}$ is the Hamiltonian of radiation in the contracting universe, which yields to the temperature $T$, i.e. $\rho_{\text{radiation}}\sim T^4\sim \rho_0a^{-4}$, where $a$ is the scale factor. The constants $\Lambda_1$ and $\Lambda_2$ are chosen to ensure a Minkowskian spacetime in the vacuum when $\rho_{\text{rad}}=0$ and both of $\phi$ and $\chi$ rest at their minima. As a toy model, let's assume that the scalar field $\chi$ as well as the self-coupling term $\lambda\phi^4$ of the Higg-like scalar field are sandwiched between two bars, i.e. they are positively defined, while the rest parts of Higss-like field, as well as the gravity and radiation, live outside of bars. The Higss self-coupling term $\lambda\phi^4$ is a switch that turns on and off the ghost. Surely this model is not the most natural one, however, as a toy model, let's accept this ad hoc setup for the time being, and see what kind of novel phenomena it will grant us.

Our Higgs field $\phi$ has a tachyonic mass, but it is stablized by the quartic term $\lambda\phi^4$.  In the thermal background, the Higgs scalar field $\phi$ receives a temperature induced mass term $\xi T^2\phi^2$ and thus its VEV takes the following value: 
\beq
\langle\phi\rangle^2=\frac{m^2-\xi T^2}{2\lambda}
\eeq
and we have assumed that $H_\chi+\lambda\phi^4-\Lambda_2>0$ yet $H_\chi-\Lambda_2<0$ during the contracting phase. 

\begin{figure}
\begin{center}
\includegraphics[width=0.42\textwidth]{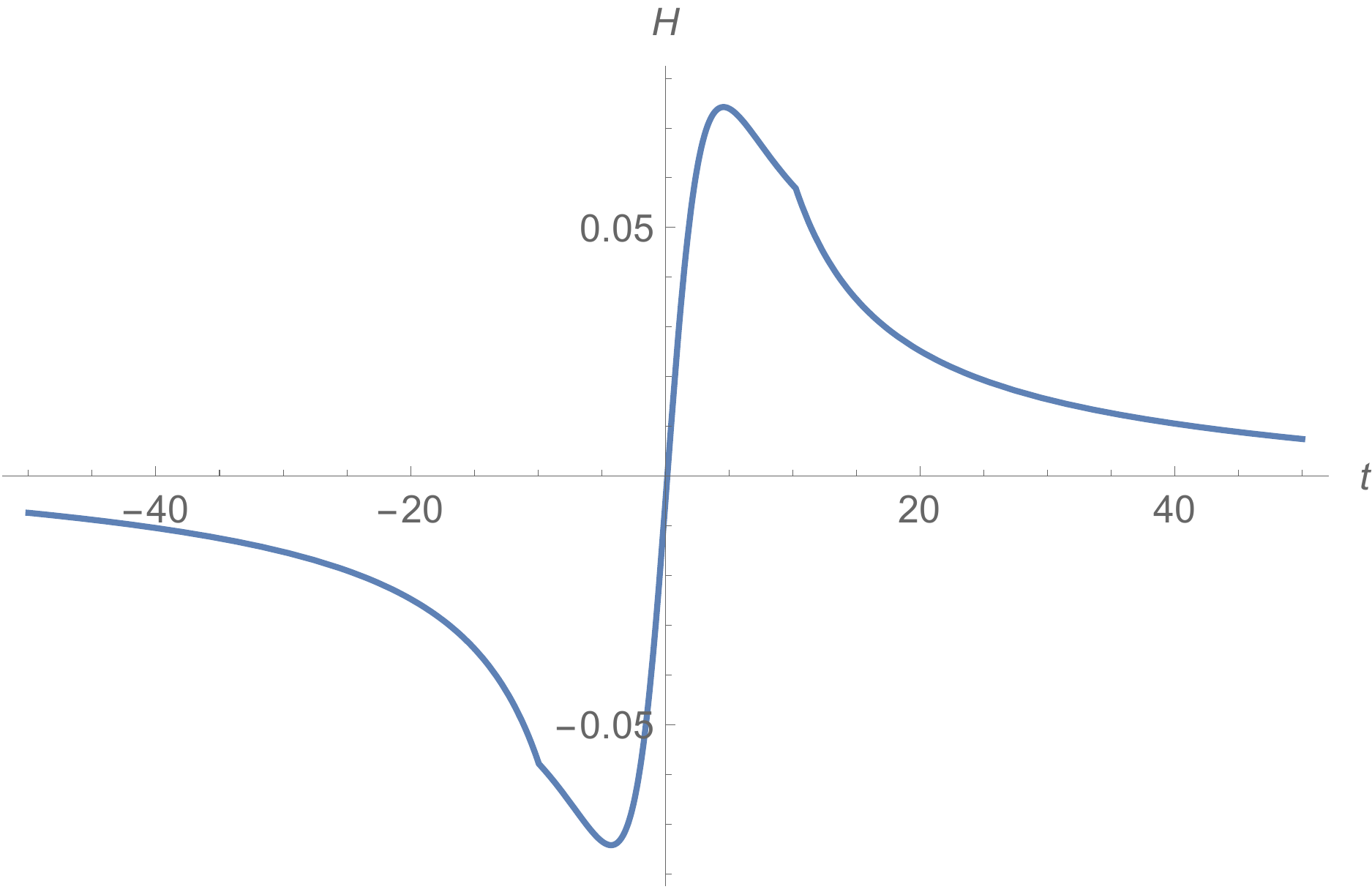}\qquad\includegraphics[width=0.38\textwidth]{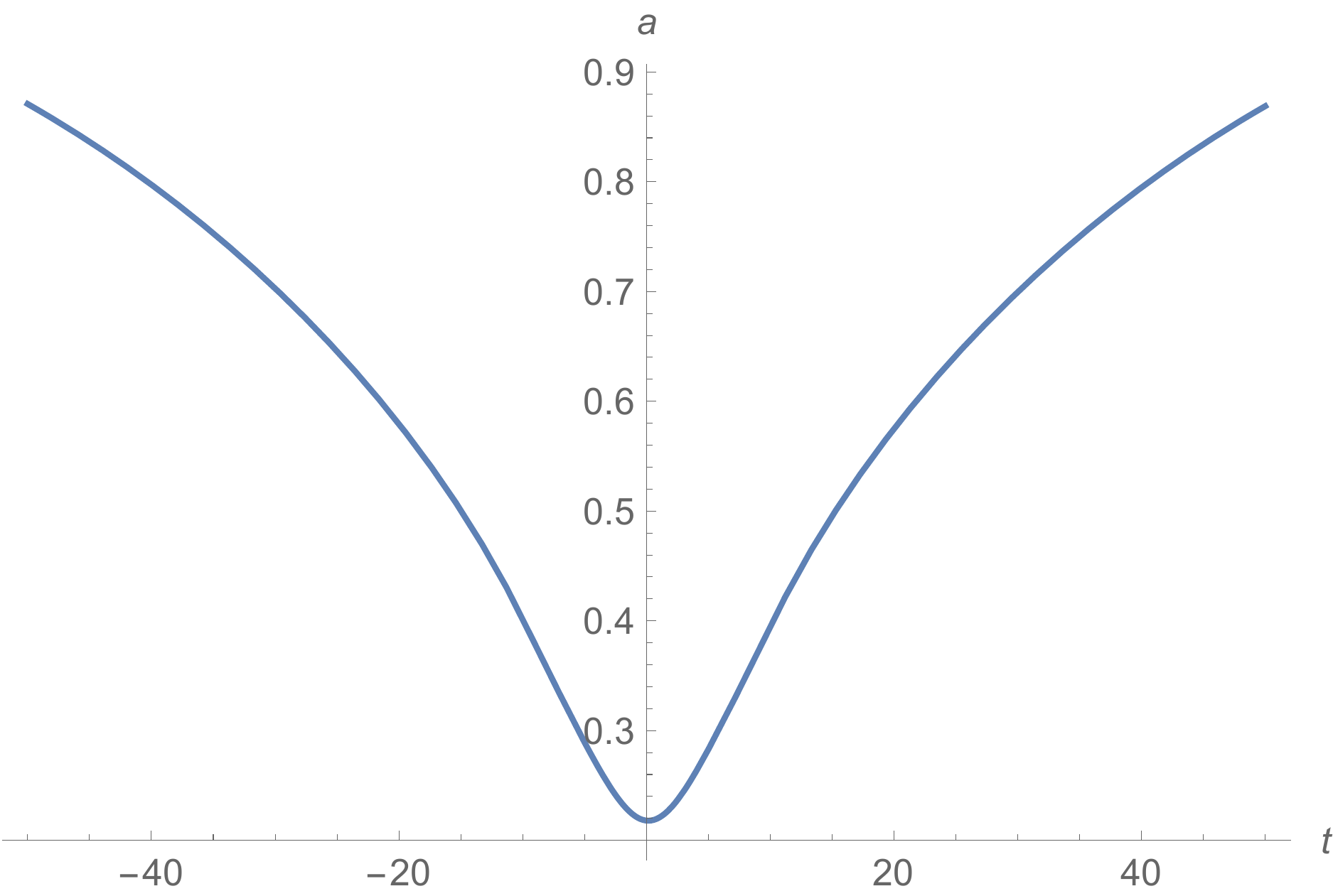}
\end{center}
\caption{The background evolution of a cosmic bounce, where the Hubble constant and time are in the Planck unit. 
}
\label{type1_vol}
\end{figure}

The temperature increases as the universe contracts, so does the thermal corrected mass term of the Higgs field. The $Z^2$ symmetry is recovered in the $\phi$ sector when the thermal corrected mass overcomes the tachyonic mass and we have $\langle\phi\rangle=0$. Since  $H_\chi-\Lambda_2<0$, the $\chi$ sector flips its overall sign and becomes a ghost. Due to the shift symmetry $\chi\to\chi+\text{constant}$, the energy density stored in the $\chi$ sector grows as $a^{-6}$ as the cosmic scale factor shrinks. Soon the $\chi$ field starts to take over universe, and we have $2\dot{H}\simeq \dot{\chi}^2$.  The null energy condition is violated and our universe undergoes a bounce phase, which is followed by big bang later on. We have numerically solved the  equations of motion, and ploted the evolution of Hubble constant and scale factor in Fig. \ref{type1_vol}.

The temperature of universe starts to cool down in the expanding phase, the Higgs field develps a non-trivial VEV again, which eventually turns off the ghost. 

\subsubsection{cosmology of type II matter coupling}
The type II matter coupling is the minimal coupling in the Jordan frame, where the Hamiltonian constraint and the gauge condition become a pair of second class constraints.  Let's begin with the theory in eq. (\ref{sqrphi}), where the gravity is canonical normalised and the matter sector is of square root form in the Einstein frame. One of the most convenient gauge fixing condition on the cosmological background is the following, 
\beq
\mathcal{G}\equiv\nabla^2\phi\approx0.
\eeq
Given proper boundary condition, the scalar field  adopts a homogeneous profile $\phi=\phi(t)$, and its fluctuation is removed by the gauge condition. The mini superspace action reads
\beq
S=\int  -3M_p^2N^{-1}a\dot{a}^2+Na^3\Lambda+a^3\sqrt{V(\phi)\left(M^4N^2-\dot{\phi}^2\right)}.
\eeq
To simplify our analysis, let's set $V(\phi)=\xi M^4$, where $\xi=\pm1$, to introduce the shift symmetry in the scalar field sector, $\phi\to \phi+\text{constant}$. Taking the variation with respect to scalar field $\phi$, we get the background equation of motion, 
\beq\label{phieom}
\partial_t\left(a^3\dot{\phi}~\sqrt{\frac{\xi}{M^4-\dot{\phi}^2}}\right)=0.
\eeq
Taking variation with respect to lapse and scale factor, we obtain
\beq
3M_p^2H^2&=&\xi M^4\sqrt{\frac{\xi M^4}{M^4-\dot{\phi}^2}}+\Lambda,\label{fr1}\\
M_p^2\dot{H}&=&-\frac{\xi}{2}\dot{\phi}^2\sqrt{\frac{\xi M^4}{M^4-\dot{\phi}^2}},\label{fr2}
\eeq
where the eq. (\ref{fr1}) is consistent with the eq. (\ref{fr2}), provided that the equation of motion for the scalar field eq. (\ref{phieom}) is satisfied. Therefore, the Bianchi identity holds. The temporal evolution crucially depends on the sign of $\xi$, and  we shall discuss the case with $\xi>0$ and the case with $\xi<0$ respectively. 
\begin{itemize}
\item $\xi>0$, and $M^4-\dot{\phi}^2>0$. The equation of motion for the scalar eq. (\ref{phieom}) implies that $\dot{\phi}\sim a^{-3}$ and it eventually rest in a vacuum state in the asymptotic future. The energy density of the scalar is diluted away as our universe expands. 
\item $\xi<0$ and thus $M^4-\dot{\phi}^2<0$. This case is more exotic and thus more interesting. Noted that eq. (\ref{fr2}) implies $\dot{H}>0$ and we have the null energy condition violation. The equation of motion for the scalar yields to the following solution, 
\beq
\dot{\phi}^2=\frac{c\cdot M^4}{c-a^6},
\eeq 
where $c$ is an intergration constant and $c-a^6>0$. As the $a^6$ approaches $c$, the $\dot{\phi}^2$ diverges, $\dot{H}\to 0^{+}$  and $H\to \text{constant}$. 
\end{itemize}
In both of cases, the Hamiltonian in both of Jordan frame and Einstein frame are positively defined and thus bounded from below, as indicated by the eq. (\ref{type2hj}) and the eq. (\ref{type2he}).

\section{Conclusion and discussion}\label{conclusion}
We  discover a general framework of the novel matter coupling in the Einstein gravity, yet the path towards this discovery is somewhat tortuous. The original motivation was to look for a theory which is as good as the Einstein gravity in the sense that all constraints are first class. However, in 4 dimensional space-time the only theory that we found is just the Einstein gravity in different guises: replacing the Hamiltonian of the Einstein gravity $H_g$ by an arbitrary function   $f(H_g)$, all constraints are still first class but the theory is  equivalent to the Einstein gravity. The equivalence is demonstrated in our current work by means of two non-perturbative approaches. The first one is the Hamiltonian analysis, where we have found that all constraints are  first class and they serve as  the space-time diffeomorphism generators; the second one is the equation of motion, where we have found that the equation of motion for graviton in a vacuum coincides with the one of GR.

 By rewriting the Einstein gravity in terms of one of its equivalents we actually define a new frame. The theoretical consistency requires the algebra must be close and the dimension of phase space must be even when matter couples to gravity in this new frame, which subjects to two types of self-consistency condition that derived in the eq. (\ref{cc}) and in the eq. (\ref{type2cond}). These self-consistency conditions impliy the Hamiltonian of both gravity and matter can have very complicated non-linear structure, which grants very rich new phenomenologies. We have worked out some classical examples, including the spherical static solution, and the non-standard FLRW cosmologies. The standard predictions are recovered at low energy scale, while new phenomenologies are granted at high energy scale. We find that the vacuum solution is just the Schwarzschild solution, as it should be since our theories are equivalent to GR. On cosmological background, the novel matter coupling may also warrants us a possibility to violate the null energy condition, without introducing the pathologies of ghost instablity. 

An interesting lesson we have learned is that sometimes the local gauge symmetries still exist, even the action is not gauge invariant.  The essence of the local gauge symmetry is that physical observables are invariant under local gauge transformations, however, the action itself is not a physical observable. In some cases, for instance the examples demonstrated in our current work, the local gauge symmetries are hidden and we have to go through all Poisson brackets to find them. 

\section*{Acknowledgments}%
We would like to thank the anonymous referee of JCAP for pointing out an error in our manuscript.  
The work of C.~L.~is carried out under
POLONEZ programme of Polish National Science Centre, 
No. UMO-2016/23/P/ST2/04240, which has received funding from the 
European Union's Horizon 2020 research and innovation programme under 
the Marie Sklodowska-Curie grant agreement No. 665778. The work of Z.L. has been partially supported by National Science Centre, Poland OPUS project 2017/27/B/ST2/02531. Authors would like to thank Katsuki Aoki, Rong-gen Cai, Yi-fu Cai, Sebastien Clesse, Francesco Di Filippo, Yohei Ema, Guilherme Franzmann, Drazen Glavan,  Bohdan Grzadkowski,  Da Huang, Qing-Guo Huang, Stefano Liberati, Hanna Lin, Shinji Mukohyama, Yun-song Piao, Jerome Quintin,  Xin Ren, Graham Ross, Kazuki Sakurai, Misao Sasaki, Henry Tye, and  Yi Wang for useful discussions.

\appendix

\section{Modified Einstein tensor}\label{MET}
The energy momentum tensor and the Einstein tensor derivations are not that straightforward if the theory is  written in terms of ADM variables, rather than metric tensor $g_{\mu\nu}$. We will discuss how to derive these tensors in this appendix. The metric tensor written in terms ADM variables reads
\beq
g^{00}=-\frac{1}{N^2},\qquad g^{0i}=\frac{N^i}{N^2},\qquad g^{ij}=h^{ij}-\frac{N^iN^j}{N^2}. 
\eeq
we have then 
\beq\label{pspn}
\frac{\partial S}{\partial N}&=&\frac{\partial S}{\partial g^{00}}\frac{\partial g^{00}}{\partial N}+2\frac{\partial S}{\partial g^{0i}}\frac{\partial g^{0i}}{\partial N}+\frac{\partial S}{\partial g^{ij}}\frac{\partial g^{ij}}{\partial N},\nonumber\\
\frac{\partial S}{\partial N^i}&=&2\frac{\partial S}{\partial g^{0j}}\frac{\partial g^{0j}}{\partial N^i}+\frac{\partial S}{\partial g^{kl}}\frac{\partial g^{kl}}{\partial N^i},\nonumber\\
\frac{\partial S}{\partial h^{ij}}&=&\frac{\partial S}{\partial g^{kl}}\frac{\partial g^{kl}}{\partial h^{ij}}.
\eeq
Reversing eqs. (\ref{pspn}) we get 
\beq\label{pspn2}
\frac{\partial S}{\partial g^{00}}&=&\frac{N^3}{2}\left[\frac{\partial S}{\partial N}+\frac{2N^i}{N}\frac{\partial S}{\partial N^i}+\frac{2N^iN^j}{N^3}\frac{\partial S}{\partial h^{ij}}\right],\nonumber\\
\frac{\partial S}{\partial g^{0i}}&=&\frac{N^2}{2}\left(\frac{\partial S}{\partial N^i}+\frac{2N^j}{N^2}\frac{\partial S}{\partial h^{ij}}\right),\nonumber\\
\frac{\partial S}{\partial g^{ij}}&=&\frac{\partial S}{\partial h^{ij}}.
\eeq

The  action of this class of GR equivalents  can be written in the first order form,
\beq
S=\int \pi^{ij}\partial_th_{ij}-N\sqrt{h}f\left(H_g\right)+2N^i\sqrt{h}\nabla_j\left(\frac{\pi^j_{~i}}{\sqrt{h}}\right),
\eeq
where
\beq
H_g\equiv R+\frac{\lambda}{h}\left(\pi^{ij}\pi_{ij}-\frac{1}{2}\pi^2\right).
\eeq
Taking the variation of the action with respect to the conjugate momentum $\pi^{ij}$, we get the relation between conjugate momentum and extrinsic curvature tensor,
\beq
\lambda\frac{\partial f}{\partial H_g}\pi_{ij}=K_{ij}-Kh_{ij}.
\eeq
According to the eq. (\ref{pspn2}), we get the modified Einstein tensor,
\beq\label{etensor1}
G^0_{~0}=\frac{1}{\sqrt{h}}\frac{\partial S}{\partial N}+\frac{N^i}{N\sqrt{h}}\frac{\partial S}{\partial N^i}
=-f+\frac{2N^i}{N}\nabla_j\left(\frac{\pi^j_{~i}}{\sqrt{h}}\right),
\eeq
\beq
G^0_{~i}&=&\frac{1}{N^2\sqrt{h}}\frac{\partial S}{\partial N^i}=\frac{2}{N^2}\nabla_j\left(\frac{\pi^j_{~i}}{\sqrt{h}}\right),\\
G^i_{~0}&=&-\frac{N^i}{\sqrt{h}}-\frac{Nh^{ij}}{\sqrt{h}}\left(\frac{\partial S}{\partial N^j}+\frac{2N^k}{N^2}\frac{\partial S}{\partial h^{jk}}\right),\nonumber\\
&=&-2N\nabla_j\left(\frac{\pi^{ij}}{\sqrt{h}}\right)+2N^j\frac{\partial f}{\partial H_g}R_{ij}-\frac{2N^j}{N}\nabla^i\nabla_j\left(N\frac{\partial f}{\partial H_g}\right)+\frac{2N^i}{N}\nabla^j\nabla_j\left(N\frac{\partial f}{\partial H_g}\right)\nonumber\\
&&+2N^i\frac{\partial f}{\partial H_g}\frac{\lambda}{h}\left(\pi^{ij}\pi_{ij}-\frac{1}{2}\pi^2\right)+2N^k\frac{\partial f}{\partial H_g}\frac{\lambda}{h}\left(-2\pi^{ij}\pi_{jk}+\pi\pi^i_{~k}\right)-\frac{2N_j}{N\sqrt{h}}\dot{\pi}^{ij},\\
G^i_{~j}&=&-\frac{N^i}{N\sqrt{h}}\frac{\partial S}{\partial N^j}-\frac{2h^{ik}}{N\sqrt{h}}\frac{\partial S}{\partial h^{kj}}\nonumber\\
&=&-2\frac{N^i}{N}\nabla_k\left(\frac{\pi^k_{~j}}{\sqrt{h}}\right)-\frac{2\dot{\pi}^{ik}h_{kj}}{N\sqrt{h}}-\delta^i_{~j}f+2\frac{\partial f}{\partial H_g}R^i_{~j}-\frac{2}{N}\left(\nabla^i\nabla_j-\delta^i_{~j}\Delta\right)\left(N\frac{\partial f}{\partial H_g}\right)\nonumber\\
&&+2\frac{\partial f}{\partial H_g}\frac{\lambda}{h}\left(\pi^{ij}\pi_{ij}-\frac{1}{2}\pi^2\right)+2\frac{\partial f}{\partial H_g}\frac{\lambda}{h}\left(-2\pi^{ik}\pi_{kj}+\pi\pi^{i}_{~j}\right)\nonumber\\
&&-\frac{2}{N\sqrt{h}}\left(\pi^{ki}\nabla_kN_j+\pi^k_{~j}\nabla_kN^i\right).
\eeq

\end{document}